\documentclass[conference]{IEEEtran}
\IEEEoverridecommandlockouts

\usepackage{cite}
\usepackage{amsmath,amssymb,amsfonts}
\usepackage{graphicx}
\usepackage{textcomp}
\usepackage{xcolor}
\usepackage{url}
\usepackage{eso-pic}

\def\BibTeX{{\rm B\kern-.05em{\sc i\kern-.025em b}\kern-.08em
    T\kern-.1667em\lower.7ex\hbox{E}\kern-.125emX}}

\begin{document}

\title{Oops, Not Now: PEARL, a RAG-Based Support Agent for Gameplay and What Players Want from AI Help}

\author{%
\IEEEauthorblockN{1\textsuperscript{st} Jiahong Li\textsuperscript{*}}
\IEEEauthorblockA{\textit{University of California Santa Cruz} \\
Santa Cruz, USA \\
jli906@ucsc.edu}
\and
\IEEEauthorblockN{2\textsuperscript{nd} Sai Siddartha Maram}
\IEEEauthorblockA{\textit{University of California Santa Cruz} \\
Santa Cruz, USA \\
msaisiddartha1@gmail.com}
\and
\IEEEauthorblockN{3\textsuperscript{rd} Atieh Kashani}
\IEEEauthorblockA{\textit{University of California Santa Cruz} \\
Santa Cruz, USA \\
atkashan@ucsc.edu}
\and
\IEEEauthorblockN{4\textsuperscript{th} Ulia Zaman}
\IEEEauthorblockA{\textit{University of California Santa Cruz} \\
Santa Cruz, USA \\
uzaman@ucsc.edu}
\and
\IEEEauthorblockN{5\textsuperscript{th} Zhiyu Lin}
\IEEEauthorblockA{\textit{University of California Santa Cruz} \\
Santa Cruz, USA \\
zlin34@ucsc.edu}
\and
\IEEEauthorblockN{6\textsuperscript{th} Cameron Marano}
\IEEEauthorblockA{\textit{University of Central Florida} \\
Orlando, USA \\
cameron.marano@ucf.edu}
\and
\IEEEauthorblockN{7\textsuperscript{th} Roger Azevedo}
\IEEEauthorblockA{\textit{University of Central Florida} \\
Orlando, USA \\
roger.azevedo@ucf.edu}
\and
\IEEEauthorblockN{8\textsuperscript{th} Jichen Zhu}
\IEEEauthorblockA{\textit{IT University of Copenhagen} \\
Copenhagen, Denmark \\
jicz@itu.dk}
\and
\IEEEauthorblockN{9\textsuperscript{th} Magy Seif El-Nasr}
\IEEEauthorblockA{\textit{University of California Santa Cruz} \\
Santa Cruz, USA \\
mseifeln@ucsc.edu}
\thanks{\textsuperscript{*}Corresponding author.}
}

\IEEEoverridecommandlockouts

\IEEEpubid{\makebox[\columnwidth]{\hfill}\hspace{\columnsep}\makebox[\columnwidth]{ }}
\AddToShipoutPictureFG*{\AtPageUpperLeft{\raisebox{-\height}{\makebox[\paperwidth]{\parbox[t]{0.86\paperwidth}{\vspace*{16pt}\centering\scriptsize Accepted for publication at the 2026 IEEE Conference on Games (CoG). \copyright~2026 IEEE. Personal use of this material is permitted. Permission from IEEE must be obtained for all other uses, in any current or future media, including reprinting/republishing this material for advertising or promotional purposes, creating new collective works, for resale or redistribution to servers or lists, or reuse of any copyrighted component of this work in other works.}}}}}
\maketitle
\IEEEpubidadjcol

\begin{abstract}
AI-powered gameplay support agents hold promise for game-based learning, yet grounding generative models in structured game data remains an open challenge. We present PEARL (Parallel Education Agent for Reflection and Learning), a dual-component Retrieval-Augmented Generation (RAG) system that combines semantic knowledge retrieval with structural board-state matching to deliver contextualized scaffolding in Parallel, a puzzle game for learning parallel programming. PEARL operates on two input streams (natural language queries and board topology), retrieving both conceptual explanations of gameplay moves and peer-generated board states as evidence: capabilities unavailable to a standard Large Language Model (LLM) with game state access alone. In a qualitative evaluation ($N{=}10$) comparing PEARL against an existing community-based Open Player Model (OPM) visualization system, participants preferred the visualization system on perceived usefulness and reported higher frustration with PEARL; five of ten minimized or abandoned the AI tool during play. Proactive delivery, generic responses, and trust deficits drove disengagement, while a subset of four participants found PEARL's grounded explanations complementary to visualization in specific contexts where they initiated the interaction. We position PEARL as a deployed design probe whose failure modes inform a concrete design agenda for AI gameplay support, captured as seven open problems for the community.
\end{abstract}


\section{Introduction}

Educational technology has increasingly embraced AI-powered tutoring systems to deliver personalized, scalable learning experiences~\cite{chen2023artificial,adiguzel2023revolutionizing}. However, a persistent tension remains: many AI tutoring systems operate as opaque pipelines whose internal reasoning is invisible to learners, limiting the metacognitive engagement that self-regulated learning theory identifies as essential for deep understanding~\cite{Winne2022metacognition,zimmerman2002becoming,Azevedo2023metacognition}.

A promising approach is \textit{scaffolding}: temporary, structured support that helps learners accomplish tasks within Vygotsky's zone of proximal development~\cite{vygotsky1978mind}, first formalized by Wood, Bruner, and Ross~\cite{wood1976role}. In technology-rich environments, scaffolding has expanded beyond one-to-one tutoring to encompass tool-mediated supports distributed across software, content, and peers~\cite{puntambekar2005tools}. Scaffolding is central to self-regulated learning~\cite{dever2023complex}, yet remains underexplored in game-based contexts where AI must interpret dynamic, structured state representations~\cite{gong2025asking}. Recent advances in conversational agents and LLM-driven game characters~\cite{huber2024leveraging,bonetti2024using,gobl2021conversational} demonstrate that generative AI can support rich, context-sensitive dialogue within games. However, while standard Large Language Models (LLMs) can generate context-sensitive responses when provided with game state information, they cannot scaffold learning through peer comparison since they lack access to verified peer play traces, community-generated board states, and expert-annotated move classifications. Without such grounding, LLMs produce generic advice and risk hallucination~\cite{lu2025novel}, a critical concern where incorrect guidance reinforces misconceptions.

This scaffolding gap is particularly acute in game-based learning environments that expose peer data through Open Player Models (OPMs). OPMs, which extend the tradition of Open Learner Models (OLMs)~\cite{Kay2022olm} to games, make player traces, inferred competencies, and peer strategies visible for self-assessment and reflection~\cite{maram2023parallel,maram2024opm}. Prior work on OPMs within Parallel, a puzzle game that teaches parallel programming through spatial coordination of concurrent processes~\cite{zhu2019programming}, has shown that peer data exposure can elicit meaningful reflection~\cite{maram2024ah,kleinman2022meaning} and improve problem-solving~\cite{kleinman2023else}. Yet even with interactive spatio-temporal visualizations~\cite{maram2024ah,kleinman2022meaning}, learners must still independently interpret what they observe, which is a non-trivial cognitive task. Prior deployments of these visualizations across approximately 50 play sessions revealed recurring struggles: trial-and-error without reflection, difficulty diagnosing failure causes, and strategy fixation~\cite{maram2024ah}. Visualization alone does not close the interpretation gap; a natural complement is \textit{conversational scaffolding} that enables players to ask questions about the peer data they are viewing.

To bridge this gap, we propose PEARL (\textbf{P}arallel \textbf{E}ducation \textbf{A}gent for \textbf{R}eflection and \textbf{L}earning), a Retrieval-Augmented Generation (RAG) system that enables pedagogical scaffolding by indexing peer play traces within a searchable vector space. PEARL employs a \textit{dual-component architecture}, combining semantic knowledge retrieval with structural board-state matching. This is because standard semantic embeddings introduce imprecision that is unacceptable for logic-based puzzles where a single misplaced edge can distinguish a correct solution from a deadlock. The semantic component retrieves conceptual explanations from an expert-annotated knowledge graph, while the structural component performs exact binary matching of board topologies to surface peer boards that are structurally proximate and pedagogically relevant. This transforms a generic conversational tutor into a peer-aware scaffolding agent that grounds every response in verified peer data.

We evaluate PEARL in Parallel~\cite{zhu2019programming}, an ideal testbed due to its customizable data pipelines, existing OPM infrastructure, and spatial board representation that mirrors techniques used in commercial games~\cite{maram2023mining}. The dual-component architecture is applicable to any game with representable state spaces and pedagogical metadata. Motivated by these considerations, we investigate the following research questions:

\textbf{RQ1:} How does PEARL's dual-component RAG architecture support gameplay assistance compared with existing visualization-based OPM support?

\textbf{RQ2:} How do players engage with PEARL's retrieval-grounded explanations and structurally retrieved peer board states during gameplay?

\textbf{RQ3:} What design considerations and player preferences emerge from deploying PEARL during live gameplay?

Based on the above research questions, we identify three core contributions:

\textbf{1. Seven empirically grounded design themes} for AI gameplay support, including on-demand availability, diagnostic failure feedback, and preservation of agency, drawn from a within-subjects qualitative deployment ($N{=}10$) and posed as a concrete design agenda for the community.

\textbf{2. A characterization of PEARL as a deployed design probe}, identifying the conditions under which retrieval-grounded help complements visualization (for a subset of participants who initiated the interaction and engaged with their current board state) and the failure modes (proactive delivery, generic responses, trust deficits) that drove a majority of participants to disengage.

\textbf{3. PEARL, a dual-component RAG system} that combines semantic knowledge retrieval with structural board-state matching, describing an architecture for grounding LLM-based agents in structured game data.

\section{Previous Work}

\subsection{Exploring Peer Learning and Peer Play}

Research has consistently demonstrated the pedagogical benefits of allowing students to explore alternative learning paths and the progress of their peers~\cite{kleinman2022meaning}. However, raw data from learning communities is often dense, multidimensional, and difficult for novices to comprehend. The field has increasingly relied on telemetry data (continuous, non-invasive logging of user behavior) to enable formative assessment, capturing the process of learning as it unfolds~\cite{gibson2016exploratory}.

These concepts extend to commercial games, where competitive titles provide post-match analytics and racing games employ ``ghosts'' for real-time peer comparison~\cite{maram2023mining}. Despite maturity in esports, such systems have rarely translated effectively into serious games~\cite{wallner2013visualization,loh2015serious}. Recent efforts bridge this gap: Maram et al.\ proposed Open Player Models (OPMs), adapting OLM principles~\cite{Kay2022olm} to game-based learning~\cite{maram2023parallel}; spatio-temporal visualizations help players interpret actions relative to alternative paths~\cite{maram2024ah,pfau2024video}; and Kleinman et al.~\cite{kleinman2023else} showed that exposing community data improves reflection quality.

Despite these advancements, a key limitation persists: even interactive visualization systems rely on the learner to independently interpret what they see. Kleinman et al.~\cite{kleinman2023else} \textit{} 
and Maram et al.~\cite{maram2024ah} both observed that participants frequently requested more explicit guidance. A natural complement is \textit{conversational scaffolding}: the process by which a more knowledgeable agent structures a task to support learner progress~\cite{wood1976role}. Scaffolding is central to intelligent tutoring systems, where VanLehn~\cite{vanlehn2011relative} showed step-level ITS approaches the effectiveness of human tutoring and dialogue-based agents such as AutoTutor~\cite{graesser2005autotutor} produce measurable learning gains~\cite{dever2023complex}. Recent work applies LLM-based scaffolding to game-based learning~\cite{gong2025asking}, though standard LLMs without grounding in verified peer traces riskhallucinated or generic advice~\cite{lu2025novel}.

\subsection{RAGs and LLMs in Education}

Large Language Models (LLMs) have demonstrated significant potential in education across language acquisition~\cite{park2024align}, humanities~\cite{vastakas2024cultural}, and STEM~\cite{achiam2023gpt}. Within CS education, LLMs aid debugging, code explanation~\cite{ma2024teach,jin2024teach}, and personalized feedback~\cite{dai2023can,sessler2025towards}.

Standard LLMs are prone to ``hallucinations,'' which Retrieval-Augmented Generation (RAG) systems address by grounding responses in verified external knowledge~\mbox{
\cite{lu2025novel,liu2025lpitutor,li2025coderag}}\hskip0pt
. Yet the prevailing interaction paradigm remains text-centric, while serious games, increasingly adopted in classrooms, embed pedagogical scaffolding within gameplay mechanics. Our paper addresses this gap by showing how RAG-based systems can interpret dynamic gameplay states without compromising pedagogical integrity.

\subsection{LLMs and RAG in Game-Based Learning}

Where earlier conversational agents relied on fixed dialogue trees (Section~II-A), recent work casts LLMs as improvisational ``gamemasters'' for role-play across civic, counseling, and historical domains~\mbox{
\cite{huber2024leveraging,bonetti2024using,stampfl2024role,maurya2024qualitative,breen2025large,kindenberg2025role}}\hskip0pt
. Despite these affordances~\cite{chen2025characterizing,sweetser2024large}, empirical evidence remains mixed: Wang et al.~\cite{wang2025effects} found LLM-adaptive mechanisms improved performance but increased cognitive load, and Tinterri et al.~\cite{tinterri2024towards} showed LLMs struggle with game-specific reasoning. RAG~\cite{lewis2020retrieval} grounds generative models in external knowledge bases.

Within gaming, RAG has primarily served \textit{narrative consistency} and \textit{memory}: the \textit{Generative Agents} architecture~\cite{park2023generative} and the ``Ghost'' system~\mbox{
\cite{zhu2023ghost} }\hskip0pt
let NPCs ``remember'' past interactions. Its use in serious games remains restricted to this textual domain, neglecting \textit{structured gameplay data}; current AI tutoring systems typically compare a student's move only against a static ``gold standard'' solution~\cite{liu2025lpitutor}.

Our work repurposes RAG for \textit{structural pedagogical retrieval}, treating dynamic player traces and board states as a queryable vector space and retrieving the trajectories of peers who navigated similar impasses, aligning with calls for ``explanatory analytics''\textit{} 
\textit{} 
~\mbox{
\cite{gibson2016exploratory} }\hskip0pt
that bridge raw log data and actionable pedagogical insight.

{
}

\subsection{Parallel -- Research Platform}

The Parallel serious game platform~\cite{maram2023parallel} facilitates the learning of parallel programming through gamified, spatial puzzles. The game presents a 2D environment where players coordinate concurrent processes (represented as moving arrows) by strategically placing and linking signals and semaphores. Prior work demonstrates the platform's versatility: Maram~\cite{maram2023parallel} developed an OPM visualization system, and Maram et al.~\mbox{
\cite{maram2024opm} }\hskip0pt
used AI to relate in-game player actions to underlying learning concepts. The spatial design of Parallel allows researchers to dissect the game board into functional ``zones,'' mirroring spatial analysis techniques used in commercial titles such as Guild Wars 2 and Dota 2. This structural similarity suggests that systems developed for Parallel possess the potential to generalize to broader gaming contexts.

{
}

\section{The PEARL System}

PEARL is a dual-component RAG \textit{} 
architecture combining semantic knowledge retrieval (Section~III-A) with structural board state matching (Section~III-B), motivated by the imprecision of standard semantic embeddings for logic-based puzzles where a single misplaced edge separates a working solution from a deadlock.

\begin{figure*}[t]
    \centering
\includegraphics[width=0.8\textwidth]{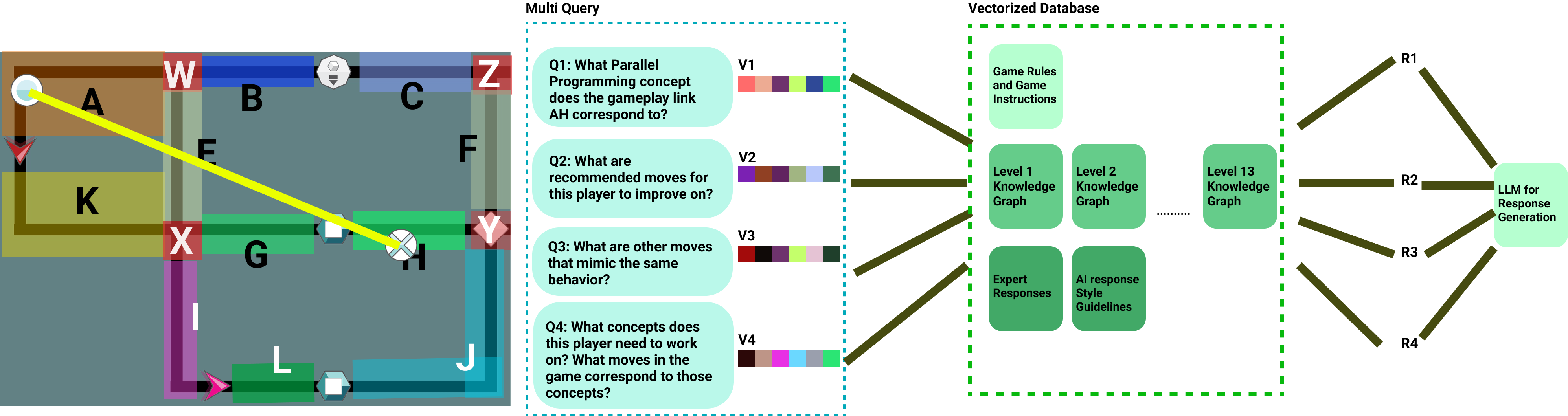}
    \caption{Semantic Knowledge Retrieval, Augmentation and Generation for Player Moves. The player's move and query are converted into a multi-query system, where queries are transformed into dense vector embeddings using GPT-embeddings. These are searched against a corpus of embedded documents from the knowledge graph, and synthesized into a unified pedagogical response.}
    \label{fig:semantic-retrieval}
\end{figure*}

\subsection{Semantic Knowledge Retrieval}

The semantic component leverages an expert-annotated knowledge graph~\mbox{
\cite{maram2024ah} }\hskip0pt
of 133 knowledge units across the two study levels. Each unit annotates a move with its classification (good or bad), the concept being demonstrated or violated, common misconceptions, and recommended alternatives. The graph encodes 391 directed recommendation edges, driving sequential chaining from semantic to structural retrieval.

\textbf{Embedding and Retrieval.} We convert this knowledge graph into a searchable vector space using GPT-embeddings (\texttt{text-embedding-3-small}, 1536-dimensional). For each move, we construct a context string that includes the link notation (the text encoding of each directed edge, e.g., \texttt{ZoneA→ZoneB}), level identifier, classification, explanation text, and board state description. As illustrated in Fig.~\ref{fig:semantic-retrieval}, \textit{} 
each player action generates a composite query integrating four sub-queries: move notation, current level context, knowledge-graph teaching content with recommended alternatives, and serialized board topology. The composite is embedded once and cosine-similarity-ranked against the indexed corpus, with a similarity floor of 0.05 and deduplication by player. The matched entries are aggregated into a ranked set of knowledge units (K\textsubscript{1}, K\textsubscript{2}, \ldots, K\textsubscript{n}), synthesized into explanatory text and a set of \textit{recommended next moves} that anchor the subsequent structural retrieval stage.

\textbf{Retrieval Parameters.} The semantic phase returns the top-$k_s = 4$ alternative good moves for correct player actions, or top-$k_s = 3$ similar mistakes for incorrect actions. The structural phase, detailed below, ranks peer boards encoded as binary adjacency vectors of length $|V|^2$ (64-dimensional on Level~3 with $|V|=8$ zones, 256-dimensional on Level~13 with $|V|=16$ zones) by Hamming distance, restricted to peer boards containing the recommendation edges surfaced by the semantic phase, and retains the top-$k_p = 5$ solution boards.

\subsection{Structural Board State Retrieval}

\begin{figure*}[t]
    \centering
\includegraphics[width=0.8\textwidth]{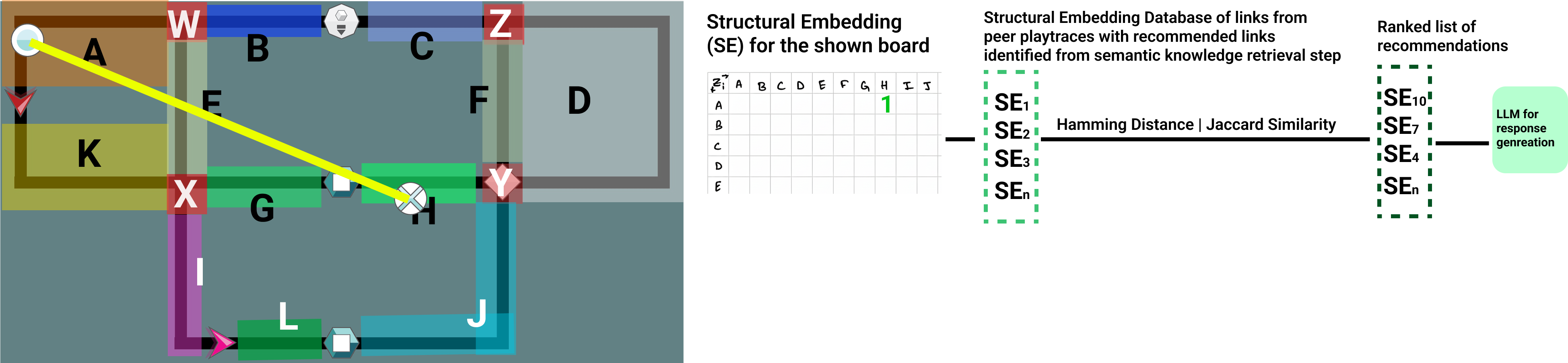}
    \caption{Structural Board State Retrieval. The player's board state is encoded as a binary adjacency vector and searched against peer playtraces using Hamming Distance and Jaccard Similarity. Pedagogically-aware ranking retrieves relevant board states based on the player's learning intent.}
    \label{fig:structural-retrieval}
\end{figure*}

While the semantic component provides conceptual explanations, the structural component addresses the need for \textit{precise structural matching} of board states. Peer board states are drawn from a curated corpus of 13 unique peer play sessions across the two study levels (4 on Level~3, 9 on Level~13) sampled from the production Parallel platform. Standard semantic embeddings create ``fuzzy'' similarity measures that can cluster board states based on superficial similarities even when they differ in critical structural properties. We illustrate the structural board state retrieval process in Fig.~\ref{fig:structural-retrieval}.

\textbf{Graph-Based Representation.} We model the game board as a directed graph $G = (V, E)$, where $V$ represents zones and $E$ represents user-created synchronization links. The board state $S_t$ at time $t$ is converted into an adjacency matrix $M$, then flattened into a binary vector $\mathbf{v}_{struct}$ of length $|V|^2$.

\textbf{Structural Similarity.} We employ Hamming Distance rather than cosine similarity because each dimension of the binary vector encodes a discrete structural property (edge present or absent), and Hamming Distance directly counts the number of structural edits between two boards. Specifically, we compute the Hamming Distance in Eq.~(1):
\begin{equation}
H(\mathbf{v}_{curr}, \mathbf{v}_{peer}) = \sum_{i=1}^{|V|^2} |\mathbf{v}_{curr}[i] - \mathbf{v}_{peer}[i]|
\end{equation}
and Jaccard Similarity for measuring overlap in Eq.~(2):
\begin{equation}
J(\mathbf{v}_{curr}, \mathbf{v}_{peer}) = \frac{|\mathbf{v}_{curr} \cap \mathbf{v}_{peer}|}{|\mathbf{v}_{curr} \cup \mathbf{v}_{peer}|}
\end{equation}

\textbf{Pedagogically-Aware Ranking.} The structural component employs a two-stage ranking process. In \textit{Validation Mode} (``Am I on the right track?''), the system minimizes Hamming Distance to retrieve structurally similar peer boards, reinforcing the player's current strategy. In \textit{Correction Mode} (``What is wrong?''), the system retrieves boards that are initially similar but diverge in key \textit{success edges} (edges present in passing peer solutions but absent from the player's current board):
\begin{equation}
Rank_{corr}(\mathbf{v}_{curr}) = \arg\min_{\mathbf{v}_{peer} \in \mathcal{S}} H(\mathbf{v}_{curr}, \mathbf{v}_{peer})
\end{equation}
where $\mathcal{S}$ is the set of peer boards that led to a correct solution. This restricts candidates to passing boards and prioritizes those closest to the player's current state, highlighting the structural ``delta'' required to reach a passing state.

\subsection{Generative Synthesis and Integration}

PEARL \textit{}
chains the components sequentially: semantic retrieval returns knowledge units (K\textsubscript{1}--K\textsubscript{n}) that the LLM synthesizes into explanatory text and recommended next moves; these recommended moves then act as a \textit{pedagogical filter} on structural retrieval, so only peer boards containing the recommended move types are eligible. The final LLM call assembles both semantic knowledge units and filtered peer board states into a single context window, generating a response that pairs explanatory text with visually rendered peer boards.

\begin{figure*}[t]
    \centering
\includegraphics[width=0.8\textwidth]{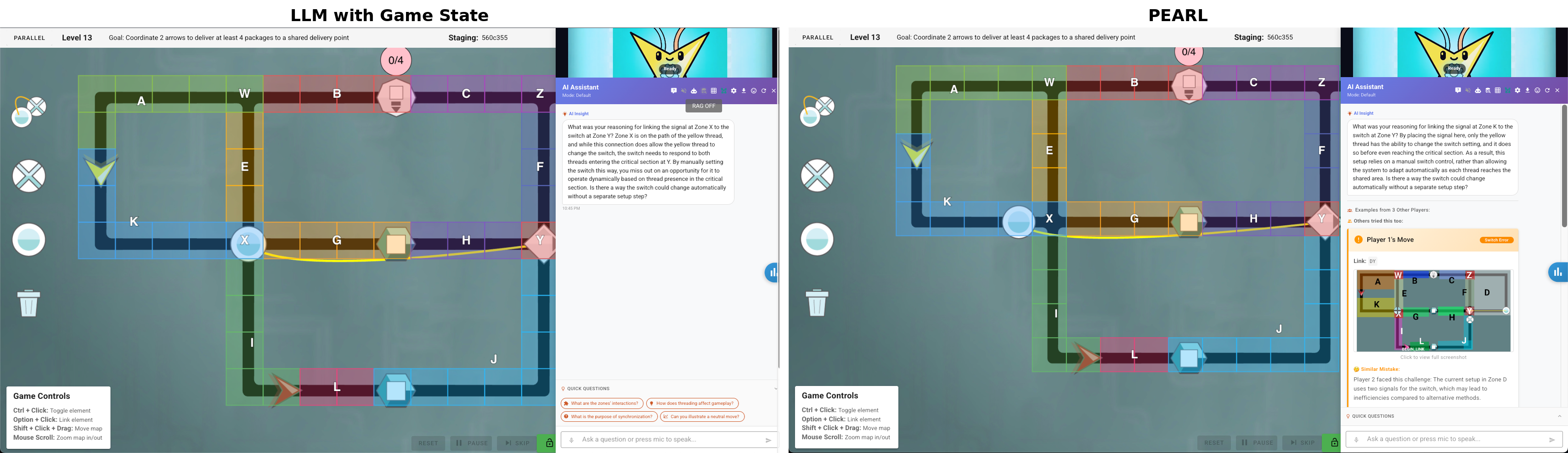}
    \caption{Side-by-side comparison of an LLM with game state (left) and the full PEARL architecture (right) for the same player state. The LLM can reason about game mechanics but lacks access to the knowledge graph and peer database. PEARL retrieves expert-annotated explanations and structurally matched peer boards to scaffold learning.}
    \label{fig:rag-comparison}
\end{figure*}

\textbf{PEARL vs.\ LLM with Game State.} Fig.~\ref{fig:rag-comparison} compares PEARL with a standard LLM that receives the player's board state, game rules, and level instructions injected into its context window. The LLM can explain parallel programming concepts and produce Socratic prompts, but cannot access the expert-annotated knowledge graph, the peer play trace database, or structural matching. The critical difference is not context awareness but \textit{scaffolding through peer evidence}, grounding abstract advice in concrete community data.

\textbf{Proactive Delivery.} When enabled (default on), PEARL fires a retrieval-and-synthesis pass on every distinct link creation event, with per-session deduplication but no temporal cooldown or move-count throttle. This delivery model drove the disengagement patterns reported in Section~\ref{sec:clippy}. Users may toggle proactivity off via a sidebar control.

\section{Evaluation}

To evaluate PEARL, we conducted a within-subjects qualitative comparison against ``Community''~\cite{maram2023parallel}, the existing OPM visualization system. Community presents peer data through spatio-temporal heatmaps, node-edge solution graphs, and aggregate performance data per level. It serves as the non-scaffolded baseline: players can \textit{see} peer data but must independently interpret it. The objective was not to establish performance superiority but to characterize interaction patterns and identify contexts where users engage with PEARL's scaffolding versus visualization-based exploration. The IRB-approved study used a think-aloud protocol with $N{=}10$ participants holding at least an undergraduate Computer Science degree.

\textbf{Study Levels.} Levels~3 and 13 were selected to span a calibrated difficulty curve while remaining solvable within 20 minutes: Level~3 introduces foundational synchronization in an 8-zone topology; Level~13 requires multi-thread coordination across 16 zones with both signal and semaphore mechanics.

\textbf{Study Design.} We employed a within-subjects design with counterbalancing across two conditions: PEARL and Community. Participants were randomly assigned to one of two counterbalanced groups: Group~1 played Level~3 with Community then Level~13 with PEARL; Group~2 played the reverse. Each session lasted 20 minutes, preceded by a tutorial on the assigned tool. After each condition, participants completed three surveys: the Technology Acceptance Model (TAM)~\cite{davis1989technology}, the NASA Task Load Index (NASA-TLX)~\cite{hart1988development}, and the Player Experience of Need Satisfaction (PENS)~\cite{ryan2006motivational}. Following both conditions, participants completed a knowledge test covering game-specific concepts (race conditions, semaphores, critical sections, and condition variables), co-designed with subject matter experts and the lead developers of Parallel. Participants then completed a semi-structured interview about their experiences with both systems.

\textbf{Analysis.} We applied codebook thematic analysis~\cite{braun2021can} with the framework method~\cite{gale2013using}. A 14-code codebook (Table~\ref{tab:codebook}) was developed deductively from the research questions. Two researchers independently coded one transcript (14 segments) with 92.9\% agreement (Cohen's $\kappa = .92$~\cite{landis1977measurement}); the codebook was then locked and remaining transcripts divided for independent coding. Coded data were charted into a participant $\times$ condition $\times$ theme matrix.

\begin{table}[t]
\caption{Codebook: 14 deductive codes organized by research question.}
\label{tab:codebook}
\centering
\footnotesize
\begin{tabular}{@{}p{0.38\columnwidth}p{0.55\columnwidth}@{}}
\hline
\textbf{Code} & \textbf{Description} \\
\hline
Interaction Patterns & How participants engaged with each tool \\
Complementary Roles & Perceived distinct functions of PEARL vs.\ Community \\
Motivation Framing & Intrinsic vs.\ extrinsic motivation shaping tool use \\
Learning \& Reflection & Conceptual learning or self-reflection episodes \\
Grounding & Responses perceived as evidence-based vs.\ generic \\
Socratic Approach & Reactions to guided questioning style \\
Basics \& Fundamentals & Naming or explaining core programming concepts \\
Analogies & Comparisons to external tools or experiences \\
Peer Board Utility & Value of structurally retrieved peer boards \\
Granularity & Desired level of detail in support \\
Autonomy Preservation & Desire to maintain independent problem-solving \\
Trust \& Credibility & Confidence in system accuracy and reliability \\
Timing of Help-Seeking & When and why participants sought help \\
AI Proactivity & Reactions to unsolicited AI suggestions \\
\hline
\end{tabular}
\end{table}

\section{Results}

Codebook thematic analysis of interview transcripts revealed three overarching patterns corresponding to our research questions; both conditions expose community peer data, the distinction being \textit{how}: Community through visualization, PEARL through retrieval-grounded scaffolding. We use pseudonyms (P1--P10) and report survey data ($n{=}10$ for TAM and NASA-TLX; $n{=}9$ for PENS due to one incomplete response) descriptively to contextualize qualitative findings.

\subsection{RQ1: Complementary but Unequally Adopted}

Most participants preferred Community over PEARL when given the choice, and at least five of ten minimized or abandoned the AI tool during play. Descriptive survey scores are consistent with this pattern: TAM Perceived Usefulness was higher for Community ($M{=}4.98$ vs.\ $M{=}4.57$, 7-point scale), as was Perceived Ease of Use ($M{=}5.02$ vs.\ $M{=}4.75$). NASA-TLX overall workload was comparable ($M{=}46.9$ vs.\ $M{=}49.1$), while the frustration subscale was higher for PEARL ($M{=}56.5$ vs.\ $M{=}43.2$).

The reasons for this preference, however, were not uniform. Some \textit{}
\textit{}
\textit{}
encountered \textit{content quality} issues: P5 found responses followed ``\textit{the same kind of script each time},'' and P1 read a stale proactive message referencing an already-removed element, eroding trust. Others rejected the \textit{interaction model}: P8 rejected hint systems outright, feeling the game was ``\textit{hand-holding me}.'' The most consistent thread of resistance, PEARL's proactive delivery, is examined in Section~\ref{sec:clippy}. P2's resistance was personality-driven, ``\textit{not always reaching out for help first.}''
\textit{}

Among participants who engaged substantively with both tools, four of ten (P4, P6, P7, P10) perceived them as serving distinct roles. P4 summarized: ``\textit{Community gives you somewhere to start\ldots AI just knew a lot, it told me exactly what it was: race conditions, semaphores, critical regions.}'' P10 drew a similar contrast: ``\textit{Community is the same for everyone\ldots the AI is based on exactly where your stuff is right now.}'' P6 preferred PEARL because ``\textit{it already knows your current solution state, you don't have to give it information},'' and P7 found it ``\textit{more helpful as a companion.}'' These participants valued PEARL's contextual awareness, its ability to ground feedback in the player's current board state, a capability Community's fixed peer screenshots could not provide.

\subsection{RQ2: Conceptual Grounding Through Dual Retrieval}

When PEARL's retrieval worked, participants engaged with the semantic and structural components in distinct and productive ways. The semantic component served a \textit{naming} function: it made implicit programming concepts explicit. P4's ``aha moment'' illustrates this: ``\textit{AI discussed mutexes, which connected to my earlier intuitive solution. My initial answer was correct but because I didn't know what mutexes were, it was confusing.}'' P6 similarly noted that AI ``\textit{clarified why zones fail}'' while trial-and-error remained the primary discovery mechanism.

However, for three of ten participants, the semantic component produced responses perceived as generic or repetitive. P5 reported the AI followed ``\textit{the same kind of script each time},'' and P8 found feedback limited to ``\textit{interesting choice you made\ldots critically think about your thing,}'' offering no actionable guidance. P3's experience captured a temporal mismatch: AI explanation eventually clarified his failure cause, but ``\textit{it wasn't necessarily helping for [the first 20 minutes]}'', suggesting retrieval latency between player action and useful feedback was too long for some.

Retrieved peer boards served two roles: \textit{validation} and \textit{alternative discovery}. Seven of ten participants reported that seeing peer struggles provided emotional reassurance (``\textit{at least I'm not the only one struggling}'', P6), while five of ten used peer boards to discover structurally different approaches. P10 compared peer boards to ``\textit{LeetCode post-solve solutions, seeing signals on the right side when mine were on the left}.'' However, P3 found peer boards ``\textit{discouraging rather than collaborative},'' and four of ten participants found the structural matching confusing when retrieved boards appeared visually dissimilar even when conceptually aligned (``\textit{visually completely off from mine but conceptually around the same}'', P4).

Scaffolding perceptions varied by interaction style. Participants who used PEARL's suggested questions perceived scaffolding (``\textit{it didn't just tell me how to optimize, it suggested zones and deduplication}'', P10), i.e., eliminating redundant links, while those who asked direct questions received more explicit answers. P4 noted that PEARL ``\textit{gives too many answers}'' on easier levels, suggesting the system lacked calibration to problem difficulty.

\subsection{RQ3: Autonomy, Timing, and Granularity as Design Axes}

Three dimensions of individual difference emerged as structurally important for AI support design.

\textbf{Autonomy preservation.} \textit{} 
\textit{}
\textit{}
\textit{}
\textit{}
\textit{}
\textit{}
\textit{}
Participants varied in how strongly they protected their independent problem-solving. The most consistent expression of this concern, resistance to unsolicited proactive help, is examined in Section~\ref{sec:clippy}.

\textbf{Help-seeking timing.} Most participants were failure-triggered, seeking help only after repeated failures. The threshold varied: P8 required ``\textit{consistent}'' wall-hitting before opening any tool, while P6 opened tools out of curiosity. P2 resisted tools entirely until time pressure forced engagement: ``\textit{part of my personality\ldots not always reaching out for help first.}''

\textbf{Desired granularity.} Preferences ranged from minimal nudges to complete explanations, often depending on proximity to a solution. P7 articulated this contingency: ``\textit{If I'm not even close\ldots I might need a full explanation. But today I was really close, I just needed a hint.}'' P6 recognized the pedagogical value of hints despite personally wanting more: ``\textit{small hints are better\ldots AI adapted better because it didn't fully give answers.}'' At the other extreme, P1 wanted ``\textit{the whole picture}'' and found the community's complete solution ``\textit{easier to deal with.}''

\subsection{Proactive Help and the Clippy Failure Mode}
\label{sec:clippy}

The single most consistent thread across participants was discomfort with PEARL's proactive intervention. Four of ten participants (P2, P3, P7, P10) cited proactive interruption as a concrete negative, and the language they used to describe it converged on a familiar trope. P3 found PEARL's proactive responses ``\textit{jarring}'', explaining that ``\textit{when I'm putting an element on the track\ldots it interrupts and tries to give advice},'' and described his preferred AI as ``\textit{very non-invasive}.'' P10 noted that the AI ``\textit{was telling me the answer before I asked it.}'' P2 found the agent ``\textit{interruptive}'' as it would ``\textit{pop up with a new message}'' after each action. Even P7, who valued PEARL overall, flagged that it ``\textit{gives you feedback every single time you make updates, even though you haven't finished your updates.}''

The objection was rarely to the \textit{content} of the help. Several of the same participants endorsed PEARL's grounded explanations when they sought them on their own terms. What participants rejected was the unsolicited delivery: an agent that responded to every move acted, in their framing, less like a tutor and more like an unwanted helper interrupting work in progress. The architectural capability for retrieval-grounded help can be undermined by the interaction model through which that help is delivered, a pattern reminiscent of \textit{Clippy} and other proactive office assistants.

\subsection{Knowledge Test}

Knowledge test scores are reported descriptively. Ten participants completed an eight-item post-session assessment covering parallel programming concepts encountered during gameplay. Mean total accuracy was 5.50 of 8 (68.8\%, $SD{=}0.94$; range 4.0--7.0), with item-level accuracy spanning 10/10 (definitions of ``semaphore'' and ``critical section'') to 4/10 (identifying intersections as the appropriate first-semaphore location). On a multi-select item identifying critical-section zones on a labeled board, no participant produced the complete correct set (2 selected only correct but incomplete zones, scored 0.5; 8 included an incorrect zone, scored 0). The within-subjects design administered the test once post-session, covering both conditions; scores reflect aggregate post-study understanding rather than condition-attributable learning gains, and $N{=}10$ precludes inferential testing of condition effects.

\section{Discussion}

The most consistent design lesson from our deployment is a pointed question: \textit{are we building Clippy?} As detailed in Section~\ref{sec:clippy}, the barrier to adoption was primarily interaction design rather than retrieval architecture: AI gameplay agents should default to passive availability and respect player-initiated interaction boundaries~\mbox{
\cite{zimmerman2002becoming}}\hskip0pt
.

Beyond proactivity, our findings position PEARL not as a replacement for visualization-based OPMs but as a \textit{conditionally complementary scaffolding modality} that addresses the interpretation gap identified in prior work~\cite{kleinman2023else,maram2024ah}. Where visualization exposes peer data, PEARL scaffolds learners' interpretation of it by grounding explanations in verified peer experiences. Participants who engaged substantively with both tools perceived them as serving different roles (concrete evidence versus personalized explanation). This complementarity, however, was asymmetric and conditional: most participants gravitated toward Community, and several disengaged from PEARL. \textit{} 
\textit{} 
Participants who received well-grounded responses (P4, P6, P7) reported productive engagement with PEARL under solicited conditions.

\textit{} 
\textit{} 

Across all ten participants, we identified seven recurring themes characterizing what players want from AI gameplay support. Three are addressed by PEARL's architecture: \textit{diagnostic failure feedback} (semantic retrieval), \textit{conceptual bridging} (knowledge graph), and \textit{visual and multimodal support} (peer board states). Two require interaction redesign: \textit{on-demand availability} and \textit{Socratic guided questioning}, valued but too generic in practice. Two remain open: \textit{preservation of agency}, requiring adaptive calibration, and \textit{genuine social community}, a desire for live peer interaction beyond retrieved boards. We pose these as a design agenda for AI gameplay support.

Our findings also suggest \textit{motivation context} shapes receptivity: participants framed gameplay as intrinsically motivated, and unsolicited help ``spoils the fun'' (P8, P3).

\textbf{Limitations and Future Work.} Our sample ($N{=}10$) of CS-trained participants limits generalizability to novice learners and non-technical domains, and the within-subjects design introduces potential order and difficulty confounds despite counterbalancing. The PEARL/Community comparison cannot isolate retrieval architecture effects from differences in modality, timing, interface, and interaction style; it characterizes two contrasting support paradigms rather than providing an architectural comparison. Some retrieval quality issues may also reflect knowledge graph coverage gaps. We evaluate PEARL as a deployed design probe rather than a retrieval system in isolation; systematic comparison of semantic-only, structural-only, and combined retrieval against expert-labeled ground truth (e.g., Precision@k, Recall@k, MRR) is left to a focused architectural study. Future work should evaluate PEARL with larger, more diverse samples, explore adaptive proactivity models that default to passive assistance, and extend the framework beyond Parallel.

\bibliographystyle{IEEEtran}
\bibliography{sample-base}

\end{document}